\documentclass[twocolumn,prl,showpacs]{revtex4}
\usepackage{amsmath}
\usepackage{amssymb}
\usepackage{epsfig}
\usepackage{bm}
\usepackage{hyperref}
\newcommand{\SSEOne}[1]{\ensuremath{|\uparrow \uparrow   \downarrow_#1 \rangle }}
\newcommand{\SSETwo}[1]{\ensuremath{ |\downarrow \uparrow  \uparrow_#1 \rangle}}
\newcommand{\SSEThree}[1]{\ensuremath{ |\uparrow \downarrow \uparrow_#1 \rangle}}
\newcommand{\VMatrix}[2]{\ensuremath{ \mathcal{V}_{#1,#2} }}
\newcommand{\HMatrixB}[2]{\ensuremath{ \mathcal{H}_{b,#1,#2} }}
\def\tr{\mathrm{Tr}}
\def\UnitMatrix{\mbox{1\hspace{-.25em}I}}

\def\GammaTilde{{\widetilde{\Gamma}}}
\def\GammaGeneric{{\Gamma^\dagger_{\alpha \beta,\tau}}}
\def\GammaSSEOne{{\Gamma^\dagger_{\uparrow \uparrow, \downarrow}}}
\def\GammaSSETwo{{\Gamma^\dagger_{\downarrow \uparrow, \uparrow}}}
\def\GammaSSEThree{{\Gamma^\dagger_{\uparrow \downarrow, \uparrow}}}
\def\HMatrixZero{{\mathcal{H}_0}}
\def\PsiTilde{{\widetilde{\Psi}^\dagger}}
\def\BEA{\begin{eqnarray}}
\def\BSES{\begin{subequations}}
\def\BE{\begin{equation}}
\def\EA{\end{array}}
\def\EEA{\end{eqnarray}}
\def\ESES{\end{subequations}}
\def\EE{\end{equation}}
\def\LP{\left(}
\def\RP{\right)}
\def\LA{\langle}
\def\RA{\rangle}
\def\dlt{\omega}
\def\rmge{\Omega}
\def\mgf{\mathcal{B}}
\def\lat{L}
\def\dis{N}
\def\hrd{H_{3\mathrm{S}}}
\def\esp{\mathbf{s}}
\def\espa{\mathbf{s}_{0}}
\def\espb{\mathbf{s}_{\dis}}
\def\sze{s^{z}}

\def\sza{S_{\mathrm{A}}^z}
\def\szb{S_{\mathrm{B}}^z}
\def\isa{\mathbf{S}_{\mathrm{A}}}
\def\isb{\mathbf{S}_{\mathrm{B}}}
\def\rji{{\tilde J}} 
\def\szt{S^z_{\mathrm{total}}}
\def\psio{|\psi_0\rangle}
\def\sta{|\!\!\uparrow \uparrow \downarrow\rangle}
\def\stb{|\!\!\uparrow \downarrow \uparrow \rangle}

\def\cnc{\mathcal{C}}
\def\cnca{\mathcal{C}_{\uparrow\uparrow}}
\def\ampa{\mathcal{A}_{\uparrow\uparrow}}
\def\cncaa{\mathcal{C}_{\uparrow\downarrow}}
\def\ampaa{\mathcal{A}_{\uparrow\downarrow}}
\def\cncp{\mathcal{C}_1}
\def\cncpp{\mathcal{C}_2}
\def\cncs{\mathcal{C}_\mathrm{s}}
\def\cnct{\mathcal{C}_\mathrm{t}}
\def\Hzero{{H_0}}
\def\hml{H}
\def\hbb{H_B}
\def\him{V}
\def\bdf{B}
\def\gge{g_{\mathrm{e}}}
\def\ggi{g_{\mathrm{I}}}
\def\sia{\sigma_\mathrm{A}}
\def\sib{\sigma_\mathrm{B}}
\def\sic{\sigma_\mathrm{e}}
\def\dns{\varrho}
\def\wsp{\mathcal{W}}
\def\wGeneric{{w_{\alpha \rightarrow \beta}}}
\def\PGeneric{{P_{\alpha \rightarrow \beta}}}
\def\PUUDU{{P_{\uparrow \uparrow \rightarrow \downarrow \uparrow}}}
\def\PDUUU{{P_{\downarrow \uparrow \rightarrow \uparrow \uparrow}}}
\begin{document}
\title{Impurity entanglement through electron scattering in a magnetic field}
\author{Alexandros Metavitsiadis}
\author{Raoul Dillenschneider}
\author{Sebastian Eggert}
\affiliation{Physics Dept.~and Res.~Center
OPTIMAS, Univ.~of Kaiserslautern, 67663 Kaiserslautern, Germany}
\date{\today}
%
%
\begin{abstract}
We study the entanglement of magnetic impurities in an environment of electrons through 
successive scattering while an external magnetic field is applied. 
We show that  the dynamics of the problem can be approximately described by a reduced model 
of three interacting spins, which reveals an intuitive view on how spins can be entangled 
by controlled electron scattering. The role of the magnetic field is rather crucial. 
Depending on the initial state configuration, the magnetic field can either increase or 
decrease the resulting entanglement but more importantly it can allow the impurities to 
be maximally entangled. 
\end{abstract}
%
%
\maketitle
%
%
The past few years have seen a large explosion of interest in the studies
of the interfaces between quantum information and many body systems.
The controlled production and detection of entangled particles is the first
step on the road towards quantum information processing \cite{nielsen2000quantum}.
A variety of methods to entangle electrons or localized spins have been proposed, based
on quite different physical mechanisms 
\cite{PhysRevB.61.R16303, PhysRevLett.84.1035, PhysRevLett.88.037901, PhysRevLett.90.166803, 
PhysRevB.63.165314, PhysRevB.65.165327, PhysRevLett.89.037901, PhysRevLett.87.277901,
PhysRevLett.91.157002, PhysRevB.69.235312, PhysRevLett.91.147901,springerlink,
PhysRevLett.96.230501, PhysRevB.74.153308, PhysRevA.81.042318,RKKY_Cho}.
One particular straight-forward application is the 
generation of entanglement between magnetic impurities or 
localized spins in mesoscopic structures 
\cite{PhysRevLett.96.230501,PhysRevB.74.153308, PhysRevA.81.042318,RKKY_Cho}, e.g.~by 
means of electron scattering.  
For the purpose of studying controlled entanglement between localized
qubits one can imagine an experimental 
realization with the help of coupled quantum dots \cite{PhysRevA.57.120},
which contain electrons that have relatively long relaxation times \cite{PhysRevB.76.035315} 
and where coherent manipulations are possible \cite{Nature.453.1043}.
Moving quantum dots generated by surface acoustic waves \cite{barnes} 
have also been observed experimentally \cite{SAW_Kataoka} and proposed
as candidates for quantum computation \cite{SAW_Shi}.
The controlled insertion of single impurity atoms with two hyperfine
states (qubits) in a bath of ultra-cold atoms has also become an active field of
research \cite{widera}.
The basic idea of an indirect coupling between localized spins is not new and
has first been considered many years ago in the context of localized nuclear or electron 
spins that can couple to conduction electrons, leading to the so--called 
Ruderman--Kittel--Kasuya--Yosida (RKKY) interaction \cite{RKKY}.
However, in order to study entanglement, an effective RKKY coupling alone is not enough as
we will show.

\begin{figure}[b!]
\begin{center}
\includegraphics[angle=0,width=0.98\columnwidth]{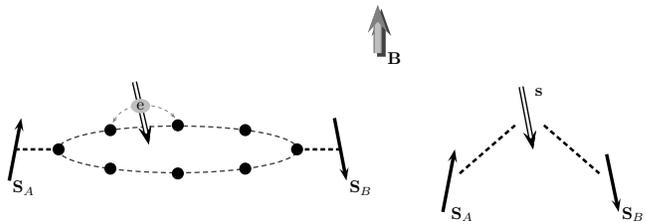}
\caption{(left) Two embedded impurities in a ring are entangled 
via successive electron scattering. In addition an external magnetic field is applied.  
(right) A minimal model of three interacting spins. 
}\label{fig-3s}
\end{center}
\end{figure}

In the present work, we consider a minimal model of two 
localized spins (qubits) that are coupled indirectly via an environment in which electrons
are injected and couple to both spins as depicted in Fig.~\ref{fig-3s} (left).
In particular, we are interested in the dynamics of the entanglement of the two qubits 
after an electron is injected into the lattice with a given initial state.  
In addition, an external magnetic field is applied on
the system  as a control mechanism over the generated entanglement.
Even though an effective RKKY interaction alone is 
not sufficient to explain the entanglement dynamics, it is sometimes 
possible to use a simplified model of only three interacting spins, Fig.~\ref{fig-3s} (right), 
exemplifying the entanglement mechanism  as well as 
the important role of the magnetic field  maximizing or destroying entanglement. 

\par 
The Hamiltonian describing this mechanism consists of 
electrons with  kinetic energy $\Hzero$, the interaction
with localized spins $\him$ and the coupling to a magnetic field $\hbb$ 
\cite{PhysRevLett.96.230501,PhysRevB.74.153308}
%
%
\BEA 
\label{EqWL1}
\hml &=& \Hzero+ \him+\hbb  \\
 & = &   \sum_{k,\sigma }\epsilon _{k}a_{k,\sigma }^{\dagger}a_{k,\sigma }
+J~(\espa\cdot\isa+\espb\cdot\isb)+  \bdf \sum_l \sze_l~.  \nonumber
\EEA
Here, $0,\dis$ are the positions of localized spins $\isa$ and $\isb$, which we will
call impurities in a system of total length $\lat$---note that there is no direct 
interaction between the impurities.  The electron spin is expressed in terms 
 of  creation  $a^\dagger$,  annihilation $a$ operators  
$\esp_l=\frac{1}{2}\sum_{\alpha,\beta} a^\dagger_{l,\alpha} 
{\bm\sigma}_{\alpha \beta} a_{l,\beta}$, with ${\bm \sigma}$ the Pauli matrices. 
Furthermore, we work in units of $J$ by setting the lattice constant $a=1$, Planck's 
constant $\hbar=1$ and Bohr's magneton $\mu_B=1$.  The hopping amplitude of 
the one--dimensional tight binding Hamiltonian is given by ${\tilde t}$, viz.~$\epsilon _{k}=-2 {\tilde t} \cos k$,  
$k=\left( 2\pi /\lat\right) n$ with  $n=-\lat/2,\ldots,\lat/2-1$. 
\par 
One of the most important symmetries of Hamiltonian \eqref{EqWL1} is the 
conservation of the total  $S^z=\sze+\sza+\szb$ component, which 
makes it possible to consider an effective field on the electrons only in Eq.~(\ref{EqWL1}).
In particular, a general magnetic field $\mgf$ applied on all 
three spins results in a Zeeman interaction of the form 
$\mgf[\gge\sze +\ggi(\sza+\szb)] = \mgf(\gge-\ggi)\sze+\ggi \mgf S^z$, 
where  $\gge$ and $\ggi$ stand for the electron and impurity $g$--factors which in 
principle are not equal. 
Since $[H,S^z]=0$ the field on the total spin $S^z$ does not influence the time evolution and 
only the effective magnetic field $B=\mgf(\gge-\ggi)$ on 
the electrons has to be considered. 
If the two $g$--factor were equal $\gge=\ggi$ the magnetic 
field would have no effect on the dynamics of the system. 
%
\par 
The generated entanglement between the impurities via the electron spin 
will acquire a time evolution according to the time evolution of the initial 
state $\psio$. Among the various quantities that one can extract information for the  entanglement
of the impurities is the concurrence \cite{PhysRevLett.80.2245,PhysRevA.63.052302}, 
a pairwise measure of entanglement. The concurrence between the two impurity spins is related to 
the reduced density matrix $\dns(t) = \tr_\mathrm{e} ~|\psi(t)\RA\LA\psi(t)|$,
where the electron's degrees of freedom are traced out.
Taking into account the symmetries of the Hamiltonian the concurrence  of the two impurities 
in the natural basis $\{|\!\uparrow \uparrow\rangle, |\!\downarrow \uparrow\rangle,
|\!\uparrow \downarrow\rangle,|\!\uparrow \uparrow\rangle\}$  reads  
\cite{PhysRevA.69.022304, PhysRevB.78.224413}
\BE\label{concurrence}
\mathcal{C}(t) = 2 |\dns_{\downarrow \uparrow,\uparrow \downarrow}(t)|.
\EE
\par 
In order to gain a basic understanding of the entanglement mechanism, it is
useful to first analyze a  reduced model of only three coupled spins in a magnetic field, 
essentially obeying the same symmetries as schematized in Fig.~\ref{fig-3s} (right) 
\BEA \label{3s}
\hrd 
&=& \rji \esp \cdot (\isa+\isb) +B\sze ~. 
\EEA
The  evolution of the initial state  $\psio=|\sia,\sib,\sic\RA$,  
with $\sigma_{\mathrm{e,A,B}}=\uparrow,\downarrow$, under Hamiltonian \eqref{3s} 
at time $t$ is given by   $|\psi(t)\RA = \exp \left(i \hrd t   \right) \psio$  and  
spans the same $S^z$ subsector with $\psio$.
In the fully polarized one--dimensional  $S^z=\pm 3/2$ subsectors, 
the impurities are in a product state 
and therefore non--entangled at all times $t\geq0$. Hence, only  the $S^z=\pm\frac{1}{2}$
subsectors are relevant for entanglement generation between the impurities. 
We only consider the sector $S^z=+\frac{1}{2}$, since the spin-flipped states have the same
time evolution for $B\to -B$.
\par
To evaluate the concurrence of the minimal three--spin model  we 
diagonalize  analytically $\hrd$  within the
$S^z=+\frac{1}{2}$ subsector. Then, we 
find the time evolution of the initial state $\psio$ and  evaluate the 
reduced density matrix $\dns(t)$ which gives the time evolution of the concurrence 
using  Eq.~\eqref{concurrence}.  
In the the sector $S^z=+\frac{1}{2}$ there are two initial 
states of interest $\psio=\sta$ and $\psio=\stb$.  The corresponding 
concurrences $\cnca(t)$ and $\cncaa(t)$  evolve according to 
\BSES\label{concurrence-3s}\BEA  
\cnca(t) &=& 
\ampa \sin^2 \dlt t  ~,  \label{concurrence-upup} \\
\cncaa(t)  &= & 
\ampaa \sqrt{\cnca^2(t)
+4\LP\cncp(t) -\frac{\rmge}{\dlt} \cncpp(t) \RP^2}~, 
 \label{concurrence-updown}
\EEA\ESES
where $\ampa=\frac{\rji^2}{2\dlt^2}$, $\ampaa=1/2$, $\rmge=\rji /4+\bdf/2 $, $\dlt=\sqrt{\rji^2/2+\rmge^2}$, 
$\cncp(t) = \sin  \rmge t \cos  \dlt t  $, and 
$\cncpp(t)=\sin     \dlt t\cos\rmge t$. 
The time dependence of the concurrence is controlled by the two frequencies $\dlt$
and $\rmge$ which are asymmetric under sign change of the magnetic field; 
hence the amplitude of $\cnca$ is also asymmetric. 
The competition of two energy scales, namely $\rji$ and $\bdf$ and the choice of
initial conditions makes 
the two directions of the magnetic field $\pm \hat{z}$ not equivalent.  
\begin{figure}[t!]
\begin{center}
\includegraphics[width=0.98\columnwidth,angle=270]{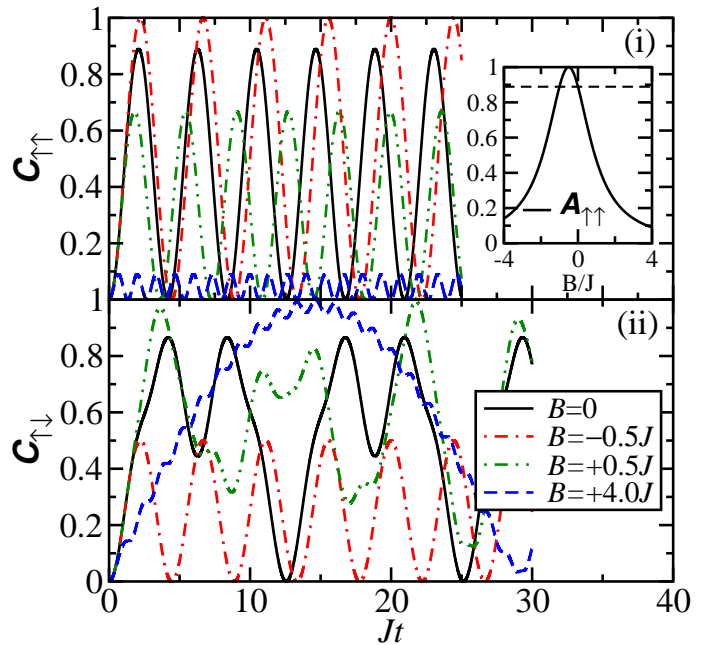}
~\\~\\
%
%
%
\caption{\label{fig-3s-updown}
(color online) Entanglement dynamics (concurrence) of the two impurities within 
model \eqref{3s} for initially aligned $\cnca(t)$ (top), 
anti--aligned $\cncaa(t)$ (bottom) impurities and different values of the applied magnetic field. 
Top inset: Amplitude $\cnca$  
versus the magnetic field.  
}
\end{center}
\end{figure}
\par 
The concurrence $\cnca(t)$ is  plotted in Fig.~\ref{fig-3s-updown}(i)  
 for different values of the applied magnetic field.  
For initially aligned  impurities, the concurrence $\cnca(t)$  simply oscillates 
with a frequency $\dlt$. In the absence of magnetic field, concurrence  
is bounded  to $8/9$ [marked with the dashed line in the inset of 
Fig.~\ref{fig-3s-updown}(i)] while when the magnetic field is switched on the
amplitude takes on the maximum value of unity  for $\rmge=0$, i.e. $\bdf=-\rji/2$. 
Furthermore, the applied magnetic field on the electron enhances entanglement
 between the impurities for $-1 <  \bdf/\rji <0$  while it reduces entanglement outside 
this range [see inset in Fig.~\ref{fig-3s-updown}(i)].  
\par 
For initially  anti-aligned impurity spins the behavior of $\cncaa(t)$ 
becomes much more interesting as shown in Fig.~\ref{fig-3s-updown}(ii).
Due to the two oscillating factors in Eq.~\eqref{concurrence-updown} 
the magnetic field allows the impurities now to be {\it almost} maximally entangled
$\cncaa\rightarrow 1$.  
For times $\dlt t_n={\pi}n$ when $\cnca$ and  $\cncpp$ vanish,  
$\cncaa(t_n)$ has a local maximum with amplitude 
\BE
\cncaa(t_n)= |\sin\frac{\rmge}{\dlt} \pi n|~,~~   n=1,2,\ldots~~ .
\EE
For increasing $n$ (larger times), one of those local maxima will become a 
global maximum with a value very close to one, see for example in 
Fig.~\ref{fig-3s-updown} the case for $B=+0.5\rji$ and $t=6\pi/\dlt\approx 21.8/\rji$. 
Note that a necessary condition for this to happen is the irrational  
value of the ratio $\rmge/\dlt$.  Moreover, for an 
irrational  value of 
$\rmge/\dlt$ the concurrence  $\cncaa(t>0)$ remains strictly larger than zero, so
the impurities are entangled for any $t>0$. 
\par
So far we have seen that $\cnca(t)$, $\cncaa(t)$ show a fundamentally 
different time evolution for any value of $\bdf$.  It is instructive to consider 
the origin of these asymmetries as a consequence of
statistical considerations. At small times, transition rates are given in terms 
$\wGeneric = \langle \alpha | \left(\UnitMatrix - i H \Delta t\right) |\beta\rangle$.
It is then straight-forward to see that for $\bdf\neq0$ the probabilities $\PGeneric = |\wGeneric|^2$
differ depending on the initial state and the direction 
of the magnetic field. For example, $\PUUDU$ will be larger than $\PDUUU$ for $\bdf>0$. 
Furthermore, for a given initial state, one can see that the concurrence will exhibit an 
asymmetric behavior with respect to the direction of the magnetic field. 
\par
Let us now consider the more realistic model depicted in Fig.~\ref{fig-3s}(left), 
obeying essentially the same symmetries as model \eqref{3s}. Here, the magnetic 
impurities $\isa$ and $\isb$ are embedded respectively at sites 
$0,\dis$ \cite{PhysRevLett.96.230501,PhysRevB.74.153308}. 
An electron is injected into the lattice,  with a given initial state,  
and scatters off the impurities.  The initial state will be of the form 
$\mbox{\SSEOne{k}}$ where the index $k$ is introduced to describe electron's 
momentum and the evolution of the initial state occurs under  
Hamiltonian \eqref{EqWL1}. Note that the electron is considered to initially occupy an  
eigenstate of $H_{0}$ with momentum $k$. 
\par 
We now focus on the time evolution of the states
$\mbox{\SSEOne{k}}$, $\mbox{\SSETwo{k}}$  and $\mbox{\SSEThree{k}}$ (spanning the subsector
$\szt=1/2$) under Hamiltonian (\ref{EqWL1}). 
To implement the time evolution of the states we consider  the time 
evolution of the spinor 
\BE \label{EqWL3}
\GammaTilde_k^\dagger = 
\left(\GammaSSEOne_k~, \GammaSSETwo_k~, \GammaSSEThree_k~\right),~~  
\GammaGeneric_k = f^\dagger_{0,\alpha} f^\dagger_{\dis,\beta}  a^\dagger_{\tau,k}~,  \nonumber 
\EE
where the fermionic operator $f^\dagger_{m,\alpha}$ creates the impurity 
on site $m$  with spin  $\alpha$. Application of the spinor $\GammaGeneric_k$
onto the vacuum generates an initial state with the corresponding parameters.  
After Laplace transforming  the time evolution we obtain 
\BE\label{EqWL4}
\left( z - \HMatrixZero \right) \GammaTilde_k^\dagger(z)
=
\GammaTilde_k^\dagger(0) 
+ \sum_q \left( \VMatrix{k}{q} + \HMatrixB{k}{q} \right) \GammaTilde_q^\dagger(z),
\EE  
where $z$ stands for the Laplace frequencies and 
the $3 \times 3$ matrices $\HMatrixZero$, $\VMatrix{k}{q}$ and 
$\HMatrixB{k}{q}$ are derived from the application 
of  $H_0$, $\him$ and $\hbb$ on the spinor 
$\GammaTilde_k^\dagger$ respectively. 
\par 
To further proceed with the solution of the system of self-consistent 
equation \eqref{EqWL4} we may apply a resonant coupling approximation 
as proposed in  Ref.~\cite{PhysRevB.74.153308}.  
That is, due to resonance, only terms with $q = \pm k$ (degenerate eigenstates of $\Hzero$)
are relevant and we end up with  a system of equations coupling only the spinors 
with wave vectors $k$ and $-k$.  This system can be reversed if we merge 
the spinors $\GammaTilde_k^\dagger(z)$  and $\GammaTilde_{-k}^\dagger(z)$ to form a new spinor $\PsiTilde_k(z)$ satisfying 
\BE \label{EqWL5}
\PsiTilde_k(z) =
\left(
\begin{tabular}{c c}
$ \wsp_{k,k}$ & 
$-\VMatrix{k}{-k} $  \\
$ -\VMatrix{-k}{k} $ & 
$\wsp_{-k,-k}$   
\end{tabular}
\right)^{-1} \PsiTilde_k(0)~, 
\EE
with $\wsp_{k,k}=z - \HMatrixZero - \VMatrix{k}{k} -\HMatrixB{k}{k} $. 
Lastly, as was shown in Ref.~\cite{PhysRevB.74.153308}, the system of equations \eqref{EqWL5}
simplifies drastically if one selects a commensurate value of the product $k\dis$, namely $k\dis=\pi\times\mathrm{integer}$. 
The time evolution of an initial state $\psio$ is obtained after inverse Laplace transform 
Eq.~\eqref{EqWL5} and the concurrence using Eq.~\eqref{concurrence} after tracing 
out electron's degrees of freedom. 
\par 
For $\psio=\mbox{\SSEOne{k}}$, we arrive at Eq.~\eqref{concurrence-upup} for $\cnca(t)$ 
with an amplitude $\ampa=(2J/L)^2/4\dlt^2$ and the two frequencies $\rmge=(2J/L)/4+B/2$, 
$\dlt=\sqrt{(2J/L)^2/2+\rmge^2}$. For $\psio=\mbox{\SSEThree{k}}$, we arrive at Eq.~\eqref{concurrence-updown} for $\cncaa(t)$ with $\ampaa=1/4$. 
Thus, at resonance,  the entanglement dynamics of the lattice is the same with the one 
of the 3--spin model with reduced  amplitudes $\ampa$ and $\ampaa$ by a factor of two 
and an effective coupling $\rji =2J/\lat$ between electron and  impurities 
driving the oscillations. The origin for the difference in the amplitude of the concurrence 
between the 3--spin model \eqref{3s} and the full model \eqref{EqWL1} arises from the 
scattering into different momentum channels for the latter. Particularly in this case,  it is 
the backscattering  that bounds the concurrence to $1/2$,  prohibiting maximum  
entanglement $\cnc\rightarrow 1$.   In fact, it is possible to consider 
a symmetric or anti-symmetric linear combination of degenerate plane 
waves with opposite momentum $k$ and $-k$.  For the symmetric (cosine) 
initial condition 
the amplitude of the concurrence is doubled resulting in maximum entanglement 
at times $\dlt t_m=\pi(m+1/2)$, 
while for the anti-symmetric linear combination no entanglement takes place.
\begin{figure}[t!]
\begin{center}
\includegraphics[angle=0,width=0.98\columnwidth]{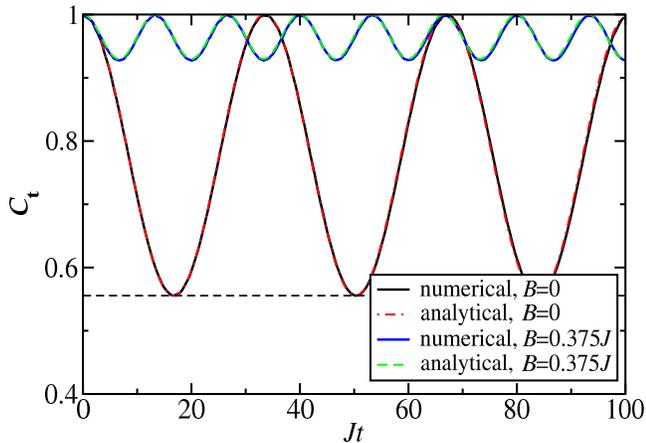}
\caption{\label{fig-triplet}
(color online) 
Entanglement dynamics (concurrence)  $\cnct(t)$ of the full model (\ref{EqWL1}) with 
$\tilde t=5J$ and $L=16$
for two values of the magnetic field $\bdf=0$ and $\bdf=0.375J$. 
The dashed line indicates the minimum 
of entanglement in the absence of magnetic field.
}
\end{center}
\end{figure}
\par 
To test our analytical calculation, we study numerically the time evolution 
of Hamiltonian \eqref{EqWL1} using exact diagonalization (ED). We consider initially 
entangled impurities in a singlet $\psio = \frac{1}{\sqrt{2}} \LP \SSETwo{k} -\SSEThree{k}\RP $ 
or triplet  $\psio = \frac{1}{\sqrt{2}} \LP \SSETwo{k} +\SSEThree{k}\RP $   combination. 
The corresponding $\cncs$ and $\cnct$ concurrence for the singlet and the triplet configuration 
are given by
\BE \label{concurrence-entangled}
 \cncs(t)=1~,\qquad\cnct(t)=1-\cnca(t)~. 
\EE
Remarkably, impurities in a singlet combination remain maximally entangled 
at all $t\geq0$. This is easily understood within the 3--spin model where the 
singlet configuration is an eigenstate of $\hrd$ leading to a time independent concurrence. 
On the other hand, the triplet initial state exhibits a sinusoidal behavior, 
shown in Fig.~\ref{fig-triplet}.  For the triplet initial 
 state, deviation from maximum entanglement occurs due to the oscillating term, the amplitude of  which 
 reduces for magnetic fields outside the range $-1<\bdf/\rji<0$, see inset 
in Fig.~\ref{fig-3s-updown}(i). Therefore, any positive or strong negative magnetic field 
applied on the electron can significantly enhance entanglement between impurities.  
\par 
In addition in Fig.~\ref{fig-triplet},  we plot numerical results obtained via ED for  
system size $\lat=16$, impurity distance $\dis=4$, and an electron in the state $k=\pi/2$. 
Numerical and analytical results 
are in very good agreement for $\cnct$ and consequently for $\cnca$ and $\cncaa$, 
 as long as the scattering of the electron on the impurities is relatively weak.
  For stronger $J$, namely for  $J/\lat$ comparable to 
the distance between the single particle energy levels, a wider range of scattering states 
participate. In this regime, the kinetic degrees of freedom become more important 
and interference of scattering into different momenta leads to non trivial entanglement dynamics. 
The resonance approximation breaks down and naturally the minimal 3--spin model fails to 
capture the entanglement process too. 
\begin{figure}[t!]
\begin{center}
\includegraphics[angle=0,width=0.98\columnwidth]{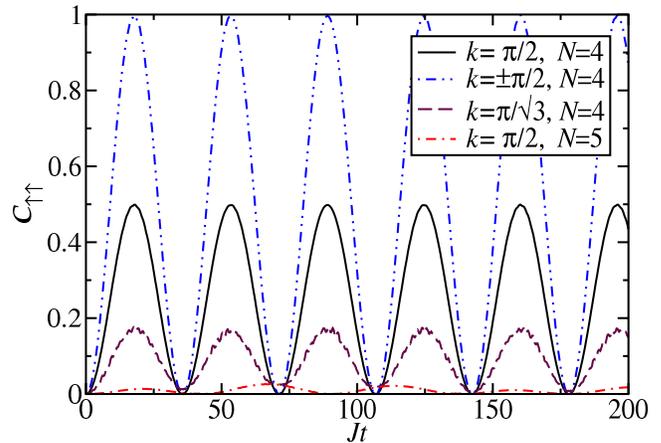}
\caption{\label{fig-numerical}
(color online) Entanglement dynamics of the full model (\ref{EqWL1}) with $\lat=16$
for two initially aligned impurities obtained via numerical
 diagonalization  for different electron initial states (see text) 
and impurity distances $\dis$. 
The magnetic field is chosen $\bdf=-J/\lat$, so that  $\rmge=0$.  }
\end{center}
\end{figure}
\par 
So far, we have focused on the generated  entanglement between the impurities 
considering different spin configurations and the influence of the magnetic field. 
Let us now work on a different scenario where the entanglement is tuned by other 
parameters like the impurity distance $\dis$ or the initial spatial part of electron's 
wave function. 
To realize this,  we work in the $S^z=1/2$ subsector for initially aligned impurities in 
a system of $\lat=16$ sites and a fixed magnetic field $B=-J/\lat$ which corresponds 
to $\rmge=0$, where maximal amplitudes are possible.  
In Fig.~\ref{fig-numerical}, we plot four characteristic examples. 
First, we plot the case where the electron is injected in the state $k=\pi/2$ and 
the impurities are $\dis=4$ sites apart. As discussed above the entanglement dynamics 
is well described by Eq.~\eqref{concurrence-upup} in this case, 
exhibiting a simple sinusoidal form with half of the maximum amplitude.  If an
electron is injected in a symmetric (cosine) superpositions of states $k=\pm \pi/2$ 
(second case)
a maximum amplitude and maximal entanglement can indeed be obtained, as also discussed above.
More interesting is a more generic case where an electron is injected into the state
$k=\pi/\sqrt{3}$, which is {\it not} an eigenstate of
$\Hzero$.  In this third case, we observe a suppression of the overall amplitude and very 
fast oscillations on top of the dominant sinusoidal form.  This pattern results from
the fact that states with different energies contribute to the
entanglement now, but nonetheless it is remarkable that the dominant sinusoidal form
remains in agreement with a simple three spin model.  The same is also true for an
incommensurate distance $N=5$ between the impurities (forth case).  The signal is now a 
suppressed superposition of oscillations with a different dominant frequency.
Nonetheless, an interpretation of such generic cases with the help of 
three spin models is in principle possible by taking the frequencies and amplitudes
as fitting parameters.
\par 
Finally, let us 
discuss the possibility of an RKKY description of Hamiltonian \eqref{EqWL1}. 
We have shown that the three-spin model in Eq.~\eqref{3s} contains 
essential features for the entanglement dynamics of the impurities. This contradicts 
the  possibility of an RKKY description of Hamiltonian \eqref{EqWL1} due to the 
different geometries of an RKKY Hamiltonian and Hamiltonian \eqref{3s}. 
Specifically, the RKKY Hamiltonian will be given by 
$H=\mathcal{J}(N)\isa\cdot \isb$ with $\mathcal{J}(N)$ some effective coupling depending 
on the distance between the impurities. The failure of such a description 
becomes  apparent if one considers the initial impurity configuration  
$|\! \uparrow \uparrow\rangle$ which is a separable state and an 
eigenstate of the RKKY interaction,  resulting to zero entanglement at all times, 
$\cnca(t)=0$. On the other hand, the configuration 
$|\! \uparrow \downarrow\rangle$, which has overlap to two 
eigenstates of the RKKY Hamiltonian, gives an oscillating behavior for the entanglement of the 
form $\cncaa(t)=|\sin \mathcal{J} t|$, which is qualitatively different from the one we 
found for model \eqref{EqWL1}.
\par 
In conclusion, we have studied the entanglement of two magnetic spin-$\frac{1}{2}$ 
impurities embedded in a tight binding ring in the presence of magnetic field. 
We showed that the main aspects of the time evolution of the entanglement 
between the impurities is revealed by a simplified model of three spins, 
Eq.~\eqref{3s}, valid for small coupling $J$ and magnetic field $\bdf$, 
disproving the possibility of an effective RKKY  description. 
Moreover, we solved analytically the full model, Eq.~\eqref{EqWL1},  
using a resonance approximation and obtained analytical formulas for various 
initial spin configurations. As far as the role of the magnetic field is concerned, 
we have shown that it can be used as a control mechanism over the generated entanglement. 
Finally, the generated entanglement is largely affected by the spatial 
part of electron's wavefunction as well as the position of the impurities in the lattice. 
\begin{acknowledgments}
This work was supported by the DFG via the Research
Center Transregio 49.
\end{acknowledgments}


\begin{thebibliography}{99}
\bibitem{nielsen2000quantum}
M. Nielsen, I. Chuang,
\emph{Quantum computation and quantum information} (Cambridge Univeristy Press, 2000).

\bibitem{PhysRevB.61.R16303}
G. Burkard, D. Loss, E.V. Sukhorukov,
Phys. Rev. B {\bf 61}, R16303 (2000).

\bibitem{PhysRevLett.84.1035}
D. Loss,  E.V. Sukhorukov,
Phys. Rev. Lett. {\bf 84}, 1035 (2000).

\bibitem{PhysRevLett.88.037901}
W.D. Oliver, F. Yamaguchi, Y. Yamamoto,
Phys. Rev. Lett. {\bf 88}, 037901 (2002).

\bibitem{PhysRevLett.90.166803}
D.S. Saraga, D. Loss,
Phys. Rev. Lett. {\bf 90}, 166803 (2003).

\bibitem{PhysRevB.63.165314}
P. Recher, E.V. Sukhorukov, D. Loss,
Phys. Rev. B {\bf 63}, 165314 (2001).

\bibitem{PhysRevB.65.165327}
P. Recher, D. Loss,
Phys. Rev. B {\bf 65}, 165327 (2002).

\bibitem{PhysRevLett.89.037901}
C. Bena, S. Vishveshwara, L. Balents, M.P.A. Fisher,
Phys. Rev. Lett. {\bf 89}, 037901 (2002).

\bibitem{PhysRevLett.87.277901}
A.T. Costa, S. Bose,
Phys. Rev. Lett. {\bf 87}, 277901 (2001).

\bibitem{PhysRevLett.91.157002}
P. Samuelsson, E.V. Sukhorukov, M. B\"uttiker,
Phys. Rev. Lett. {\bf 91}, 157002 (2003);
Phys. Rev. Lett. {\bf 92}, 026805 (2004).

\bibitem{PhysRevB.69.235312}
A.V. Lebedev, G. Blatter, C.W.J. Beenakker, G.B. Lesovik,
Phys. Rev. B {\bf 69}, 235312 (2004).

\bibitem{PhysRevLett.91.147901}
C.W.J. Beenakker, C. Emary, M. Kindermann, J.L. vanVelsen,
Phys. Rev. Lett. {\bf 91}, 147901 (2003).

\bibitem{springerlink}
G. Lesovik, T. Martin, G. Blatter,
Eur. Phys. J. B {\bf 24}, 287 (2001).

\bibitem{PhysRevLett.96.230501}
A.T. Costa, S. Bose, Y. Omar,
Phys. Rev. Lett. {\bf 96}, 230501 (2006).

\bibitem{PhysRevB.74.153308}
G.L. Giorgi, F. dePasquale,
Phys. Rev. B {\bf 74}, 153308 (2006).


\bibitem{PhysRevA.81.042318}
F. Ciccarello, S. Bose, M. Zarcone,
Phys. Rev. A {\bf 81}, 042318 (2010).

\bibitem{RKKY_Cho} 
S.Y. Cho, R.H. McKenzie,
Phys. Rev. A {\bf 73}, 012109 (2006).


\bibitem{PhysRevA.57.120}
D. Loss, D.P. DiVincenzo,
Phys. Rev. A {\bf 57}, 120 (1998).

\bibitem{PhysRevB.76.035315}
J.M. Taylor, J.R. Petta, A.C. Johnson, A. Yacoby,
C.M. Marcus, M.D. Lukin,
Phys. Rev. B {\bf 76}, 035315 (2007).

\bibitem{Nature.453.1043}
R. Hanson, D.D. Awschalom,
Nature {\bf 453}, 1043 (2008).

\bibitem{barnes} C.H.W. Barnes, J.M. Shilton, A.M. Robinson,
Phys.Rev. B {\bf 62}, 8410 (2000).

\bibitem{SAW_Kataoka} 
M. Kataoka, M.R. Astley, A.L. Thorn,
D.K.L. Oi, C.H.W. Barnes, C.J.B. Ford, D. Anderson,
G.A.C. Jones, I. Farrer, D.A. Ritchie, M. Pepper,
Phys. Rev. Lett. {\bf 102}, 156801 (2009).

\bibitem{SAW_Shi} X. Shi, M. Zhang, L.F. Wei,
Phys. Rev. A {\bf 84}, 062310 (2011).

\bibitem{widera} N. Spethmann, F. Kindermann, S. John, C. Weber, D. Meschede, and A. Widera
Phys. Rev. Lett. {\bf 109}, 235301 (2012).

\bibitem{RKKY} M.A. Ruderman and C. Kittel, Phys. Rev. {\bf 96}, 99 (1954);
T. Kasuya, Prog. Theor. Phys. {\bf 16}, 45 (1956);
K. Yosida, Phys. Rev. {\bf 106}, 893 (1957).

\bibitem{PhysRevLett.80.2245}
W.K. Wootters, Phys. Rev. Lett. {\bf 80}, 2245 (1998).

\bibitem{PhysRevA.63.052302}
K.M. O'Connor, W.K. Wootters,
Phys. Rev. A {\bf 63}, 052302 (2001).

\bibitem{PhysRevA.69.022304}
L.~Amico, A.~Osterloh, F.~Plastina, R.~Fazio,  and G.~M.~Palma
Phys.~Rev.~A {\bf 69}, 022304 (2004)

\bibitem{PhysRevB.78.224413}
R.~Dillenschneider, Phys.~Rev.~B {\bf 78}, 224413 (2008) 

%

%

\end{thebibliography}
\end{document}